\documentclass[%
 reprint,
superscriptaddress,
 amsmath,
 amssymb,
 aps,
a4,
prc,
]{revtex4-2}

\usepackage{hyperref}
\usepackage{amsmath}
\usepackage{amssymb}
\usepackage{graphicx}
\usepackage{mathrsfs}
\usepackage{color}
\usepackage{bm}
\usepackage{ulem}

\DeclareRobustCommand{\erase}{\bgroup\markoverwith{\textcolor{red}{\rule[.5ex]{2pt}{1.pt}}}\ULon}

\renewcommand{\ll}{\Lambda\Lambda}

\begin{document}

\title{Constraining the $\Lambda\Lambda$ interaction with terrestrial and astronomical data}

\author{Yusuke Tanimura}
\affiliation{Department of Physics and Origin of Matter and Evolution of Galaxies (OMEG) Institute, Soongsil University, Seoul 06978, Korea}
\author{Chang Ho Hyun}
\affiliation{Department of Physics Education, Daegu University, Gyeongsan 38453, Korea}
\author{Myung-Ki Cheoun}
\affiliation{Department of Physics and Origin of Matter and Evolution of Galaxies (OMEG) Institute, Soongsil University, Seoul 06978, Korea}

% \author{Emiko Hiyama}
% \affiliation{Department of Physics, Tohoku University, Sendai 980-8578, Japan}
% \affiliation{RIKEN Nishina Center, Wako 351-0198, Japan}

\date{\today}

\begin{abstract}
Terrestrial double-$\Lambda$ hypernuclear data and astronomical observations of neutron stars provide complementary constraints on the $\Lambda\Lambda$ interaction.
In this work, we investigate the $\Lambda\Lambda$ interaction within a 
Skyrme energy density functional framework based on the KIDS (Korea-IBS-Daegu-SKKU) models.
We employ a Skyrme-type $\Lambda\Lambda$ interaction that 
includes the standard $s$- and $p$-wave terms, as well as  
a density-dependent term that effectively represents an $N\Lambda\Lambda$ three-body force. 
The $s$-wave terms are constrained using data on double-$\Lambda$ hypernuclei 
supplemented by pseudodata obtained from core + $2\Lambda$ three-body model calculations including heavier hypernuclei. 
We show that the data on heavier systems are essential to simultaneously constrain the 
two $s$-wave parameters.
We further explore the impact of the $p$-wave and $N\Lambda\Lambda$ components on the neutron-star properties 
and find that appropriate repulsive contributions of these terms yield consistency
with current neutron-star mass-radius observations.
These results indicate that the present framework provides phenomenologically acceptable equations of state for dense $(N,\Lambda)$ matter over a wide range of densities and highlight the importance of future experimental data on heavier double-$\Lambda$ hypernuclei.
\end{abstract}


\maketitle

\section{Introduction}

A standard scenario for the diversity of elements in the universe is
that nuclei heavier than iron are created through the synthesis of radioactive isotopes 
far from the valley of stability and their decays in various modes.
The synthesis of unstable nuclei is likely to occur in environments of 
extremely high density and temperature, such as
supernova explosions and neutron-star mergers.
Similarly, exotic particles and states that are absent in normal matter may appear at extreme conditions,
and they can also play a critical role to form the present state of matter and universe.
Hyperon is a well known example.
Their production has been observed in relativistic heavy ion collisions \cite{STAR05,ALICE14,ALICE16,ALICE25,ALICE26,Fe24},
and their presence in neutron-star cores, though not directly measurable, is difficult to ignore 
in the current theoretical framework \cite{Mo04,Lim15,ChVi16,OHKT17,SHHL26,Glendenning,Lo21,SPRV06,ScRi11,ChVi16,MCS13,MCKS25,Togashi16,Yamamoto17,Looee26}.
Accurate description of their physical properties and interactions with other particles will lead to deeper and wider
understanding of the matter and the universe.

The tables compiled by the Particle Data Group \cite{pdg} show that the basic properties of the lightest hyperons in the baryon octet
are measured at high precision.
Their interactions with the nucleon are relatively well understood by means of numerous experimental data
and theoretical studies within various microscopic approaches \cite{HaTa06,FeNa15,BoBrGi12,GaHuMi16,GiHu95,Petschauer20}.
Interactions between hyperons, on the other hand, are poorly constrained.
In Ref.~\cite{La98}, interactions between $\Lambda$ hyperons (hereafter referred to as $\ll$ interaction) were 
described in the form of Skyrme force that contains five unknown parameters.
At that time the only experimental information available to constrain the $\ll$ interaction was the binding energy of $^{13}_{\ll}$B.
The two key parameters [$\lambda_0$ and $\lambda_1$ in our notation in Eq. \eqref{eq:vsk_LL}] were fit to the this datum, while the remaining three parameters ($\lambda_2$, $\lambda_3$ and $\alpha$) were set to zero.
Resulting interaction had large uncertainty and consequent neutron star equation of state (EoS)
at high densities ranged from soft to stiff regions.

Almost thirty years have passed since Ref.~\cite{La98} was published.
In the meantime, there were progress in the experiment for the double $\Lambda$ hypernuclei.
The number of data has increased and their accuracy has been improved.
In this work, we examine to what extent the $\ll$ interactions can be constrained 
by using the hypernuclear data and neutron-star properties obtained from the astronomical measurements.

There have been numerous theoretical efforts to investigate 
the effects of the $\Lambda\Lambda$ interactions, particularly in the context of 
the hyperon puzzle associated with $\approx 2M_\odot$ neutron stars.
Such studies have been carried out within Skyrme-type frameworks~\cite{Mo04,Lim15,ChVi16,OHKT17,SHHL26} 
as well as other theoretical models~\cite{Glendenning,Lo21,SPRV06,ScRi11,ChVi16,OHKT17,MCS13,MCKS25,Togashi16,Yamamoto17,Looee26}. 
The $\Lambda\Lambda$ force has also been used in studies of multi-$\Lambda$ hypernuclei~\cite{BIM,MIB83,IBM85,Rufa90,MZ93,Sch94,HKYMR10,MiHa12,MCH09,MiCh11,KMGR15,MKG17,GBKM18,Ta19,Guo22,Su24,LCZR24,ZhHiSa25}.

Very recently, an attempt has been made in Ref.~\cite{SHHL26} to constrain 
the parameters of the Skyrme-type $\Lambda\Lambda$ interaction based on the data from 
double-$\Lambda$ hypernuclei and astronomical observations. 
By systematically employing a Bayesian framework, the authors provided simultaneous constraints 
on the five parameters including those associated with effective $\Lambda\Lambda$ and $N\Lambda\Lambda$ interactions.

In this work, we adopt a complementary approach. 
While we also make use of both hypernuclear data and neutron-star observations, 
the former is supplemented by pseudodata obtained from core + $2\Lambda$ three-body model calculations for heavier systems.
The three-body model is constructed to be consistent with the existing data for single-$\Lambda$ hypernuclei and for the double-$\Lambda$ hypernucleus $^{~~6}_{\Lambda\Lambda}$He~\cite{Ahn13,Takahashi01}.
This framework allows us to separate the sensitivities of the parameters to different datasets: 
the $s$-wave parameters are well constrained by the hypernuclear information, whereas the remaining parameters are explored using the neutron-star observables. 

We find that the $s$-wave Skyrme parameters $\lambda_0$ and $\lambda_1$ cannot be determined uniquely
with the three data of binding energies for $^{11}_{\ll}$Be, $^9_{\ll}$Li and $^9_{\ll}$Be.
When four pseudodata for $^{18}_{\ll}$O, $^{34}_{\ll}$S, $^{42}_{\ll}$Ca and $^{58}_{\ll}$Ni
are included, these parameters become significantly better constrained 
within a narrow region.
However, the seven data are insufficient to constrain the density-dependence parameter $\lambda_3$, so the fitting is not
able to determine even its sign.
To further constrain $\lambda_3$ as well as the $p$-wave parameter $\lambda_2$, 
we examine their impact on the neutron-star maximum mass
and mass-radius relation.
%on the values of $\lambda_2$ and $\lambda_3$.
Although the resulting neutron-star properties exhibit some dependence on the 
underlying symmetry energy, the combination of medium-mass
double-$\Lambda$ hypernuclear information and neutron-star properties 
provides meaningful constraints on these parameters.

The paper is organized as follows.
In Sec. \ref{sec:Model}, we describe the energy density functional for $\ll$ interaction adopted in this work.
The results and discussions are presented in Sec. \ref{sec:Results}.
Section \ref{sec:Conclusions} summarizes the present study and provides an outlook.
Details of the three-body model are given in the Apprendix \ref{app:3BModel}.

\section{Model: Skyrme energy density functional}\label{sec:Model}

The KIDS model \cite{GHHY22,CHHC22} is a Skyrme-type energy density functional (EDF) for multi-$\Lambda$ hypernuclear systems, written as 
\begin{align}
E &= 
\int d^3r\ {\cal E}, 
\end{align}
where the total energy density is decomposed into three contributions,  
\begin{align}
{\cal E} &= {\cal E}_{NN} + {\cal E}_{N\Lambda} + {\cal E}_{\Lambda\Lambda}. 
\end{align}
Here, ${\cal E}_{NN}$ is the nucleonic energy density~\cite{GHHY22}, 
${\cal E}_{N\Lambda}$ includes the kinetic energy density of $\Lambda$ hyperons 
and the effective $N\Lambda$ interaction~\cite{CHHC22}, and 
${\cal E}_{\Lambda\Lambda}$ represents the $\Lambda\Lambda$ interaction. 

\if0
The $N\Lambda$ part ${\cal E}_{N\Lambda}$ is given by~\cite{CHHC22} 
\begin{align}
{\cal E}_{N\Lambda} &= 
\frac{\hbar^2}{2m_\Lambda}\tau_\Lambda
+
u_0\left( 1+\frac{y_0}{2} \right)\rho_N\rho_\Lambda 
\nonumber
\\
&\hspace{.4cm}
+
\frac{1}{4}(u_1+u_2)\left( \rho_N\tau_\Lambda + \tau_N\rho_\Lambda\right)
\nonumber
\\
&\hspace{.4cm}
-
\frac{1}{8}\left( 3u_1 - u_2 \right)\rho_\Lambda\bm\nabla^2\rho_N
\nonumber
\\
&\hspace{.4cm}
+\sum_{n=1}^N
\frac{3}{8}u_{3n}\left( 1+\frac{y_{3n}}{2} \right)\rho_N^{1+n/3}\rho_\Lambda, 
\label{eq:edens_NL}
\end{align}
where $\rho_\Lambda$ and $\tau_\Lambda$ are the number density and the kinetic-energy density 
of $\Lambda$ hyperons, respectively, and $\rho_N$ is the total nucleon density. 
\fi

The $\Lambda\Lambda$ contribution adopted in this work is given by 
\begin{align}
{\cal E}_{\Lambda\Lambda} &= 
  \frac{1}{4}\lambda_0\rho_\Lambda^2
+ \frac{1}{8}\left( \lambda_1+3\lambda_2 \right)\rho_\Lambda\tau_\Lambda
\nonumber
\\
&\hspace{.4cm}
+ \frac{3}{32}\left( \lambda_2-\lambda_1 \right)\rho_\Lambda\bm\nabla^2\rho_\Lambda
+ \frac{1}{4}\lambda_3\rho_\Lambda^2\rho_N, 
\label{eq:edens_LL}
\end{align}
where $\rho_\Lambda$ and $\tau_\Lambda$ are the number density and the kinetic-energy density 
of $\Lambda$ hyperons, respectively, and $\rho_N$ is the total nucleon density. 
This EDF is based on a Skyrme-type effective $\Lambda\Lambda$ interaction of Ref.~\cite{La98}: 
\begin{align}
v_{\Lambda\Lambda} &= 
\left[ \lambda_0 + \lambda_3\rho_N^\alpha\left( \frac{\bm r_1+\bm r_2}{2} \right) \right]\delta(\bm r_1-\bm r_2) 
\nonumber
\\
&\hspace{.4cm}
-\frac{\lambda_1}{8}
\left[ \overleftarrow{\bm\nabla}^2\delta(\bm r_1-\bm r_2)
+ \delta(\bm r_1-\bm r_2)\overrightarrow{\bm\nabla}^2\right]
\nonumber
\\
&\hspace{.4cm}
+\frac{\lambda_2}{4}
\overleftarrow{\bm\nabla}\cdot\delta(\bm r_1-\bm r_2)
\overrightarrow{\bm\nabla}, 
\label{eq:vsk_LL}
\end{align}
with $\alpha=1$. 
Here, $\overrightarrow{\bm\nabla} = \bm\nabla_1-\bm\nabla_2$ acts on the right, while 
$\overleftarrow{\bm\nabla} = \bm\nabla_1-\bm\nabla_2$ acts on the left. 
Given the limited amount of information available on the $\Lambda\Lambda$ interaction,
we fix $\alpha=1$ to reduce the number of free parameters.
The remaining coefficients $\lambda_i$ ($i=0-3$) are treated as adjustable parameters.
The terms with $\lambda_0$ and $\lambda_1$ are the $s$-wave interactions, while 
the $\lambda_2$ term is the $p$-wave one~\cite{La98,VB72}. 
The density-dependent $\lambda_3$ term introduces an effective repulsive $N\Lambda\Lambda$ three-body force. 

All the Hartree-Fock calculations with the KIDS functional in the present work 
are done by imposing the spherical symmetry and the equal-filling approximation. % unless stated otherwise. 

\section{Results and discussions}\label{sec:Results}

\subsection{Double-$\Lambda$ hypernuclei: fitting of $\lambda_0$ and $\lambda_1$}

In the present work, we employ the KIDS functionals (KIDS0, KIDS-A, KIDS-B, KIDS-C, and KIDS-D) 
for the $NN$ sector~\cite{GHHY22}. For the $N\Lambda$ sector we employ the Y4 parameter sets~\cite{CHHC22} corresponding to each $NN$ parameter set. 
The remaining $\Lambda\Lambda$ interaction parameters should then be adjusted to 
available experimental data. 
However, the data on double-$\Lambda$ hypernuclei are limited to light systems 
and insufficient to fully constrain all the parameters in Eq.~\eqref{eq:edens_LL}. 

Therefore, in the first step, we try to constrain only $\lambda_0$ and $\lambda_1$ of the $s$-wave terms. 
The $p$-wave interaction terms proportional to $\lambda_2$ are excluded because their contribution 
vanishes identically for the ground states of double-$\Lambda$ systems, 
in which the two $\Lambda$ particles occupy the $1s$ orbital~\cite{La98}. 
The density-dependent term involving $\lambda_3$ is left unconstrained as it is sensitive mainly to 
variation of strange matter properties as a function of the density. 
A discussion of the resulting high-density equation of state will be presented later.

Even the simultaneous determination of $\lambda_0$ and $\lambda_1$ is challenging 
when only very light double-$\Lambda$ hypernuclei are considered.
The parameter $\lambda_0$ mainly reflects bulk nuclear properties, 
whereas $\lambda_1$ is more sensitive to surface effects.
Therefore, a reliable determination of both parameters ideally requires data 
covering systems with different mass numbers, since the balance between 
bulk and surface contributions varies with system size.
To this end, we supplement the limited experimental information with pseudodata 
for heavier double-$\Lambda$ hypernuclei obtained from a core + $2\Lambda$ three-body model, 
which is constructed to be consistent with the existing data for single-$\Lambda$ hypernuclei
and $^{~~6}_{\Lambda\Lambda}$He (see Appendix \ref{app:3BModel} for details).

The parameters $\lambda_0$ and $\lambda_1$ will be determined 
by fitting to the experimental binding energies of double-$\Lambda$ hypernuclei. 
The single- and double-$\Lambda$ binding energies of hypernuclei are defined as 
\begin{align}
B_\Lambda\left(^A_\Lambda Z\right) &= B\left(^A_\Lambda Z\right) - B\left(^{A-1}Z\right),
\\
B_{\Lambda\Lambda}\left(^A_{\Lambda\Lambda} Z\right) &= B\left(^A_{\Lambda\Lambda} Z\right)
 - B\left(^{A-2} Z\right),
\end{align}
respectively, where $B$ denotes the total binding energy of a nucleus. 
We also introduce the two-$\Lambda$ correlation energy: 
\begin{align}
\Delta B_{\Lambda\Lambda}\left( ^A_{\Lambda\Lambda} Z \right) &= 
B_{\Lambda\Lambda}\left( ^A_{\Lambda\Lambda} Z \right) - 2B_{\Lambda}\left( ^{A-1}_{\Lambda} Z \right) , 
\end{align}
which is predominantly sensitive to the $\Lambda\Lambda$ interaction. 

The model parameters are optimized by minimizing the mean absolute relative deviation (MD),
\begin{align}
{\rm MD} &= \frac{1}{N_{\rm data}}
\sum_{i=1}^{N_{\rm data}}\left|\frac{O_{i}^{\rm calc}-O_{i}^{\rm expt}}{O_{i}^{\rm expt}}\right|, 
\label{eq:MD}
\end{align}
where $N_{\rm data}$ is the number of data points, $O_i^{\rm expt}$ and $O_i^{\rm calc}$ 
are the experimental and calculated values of the observable $i$, respectively.

Table~\ref{tb:dataset} summarizes the dataset employed for the fitting of the parameters 
$\lambda_0$ and $\lambda_1$. 
We use an experimental value of $B_{\Lambda\Lambda}$ for $^{11}_{\Lambda\Lambda}$Be reported in 
Ref.~\cite{Ek19}, 
the calculated $B_{\Lambda\Lambda}$ values of $^{~9}_{\Lambda\Lambda}$Li and $^{~9}_{\Lambda\Lambda}$Be 
from Ref.~\cite{Hi02}, and four more pseudodata for $\Delta B_{\Lambda\Lambda}$ of 
$^{~18}_{\Lambda\Lambda}$O, $^{~34}_{\Lambda\Lambda}$S, $^{~42}_{\Lambda\Lambda}$Ca, and 
$^{~58}_{\Lambda\Lambda}$Ni. The pseudodata for $A\geq 18$ are obtained from the three-body model calculations performed in this work.

\begin{table}
\caption{
The dataset used to fit the $\Lambda\Lambda$ interaction parameters $\lambda_0$ and $\lambda_1$. 
The values shown in boldface are included in the dataset, and the pseudodata are marked with an asterisk. 
The value $B_{\Lambda\Lambda}$ for $^{11}_{\Lambda\Lambda}$Be is experimental~\cite{Ek19}, 
whereas those for $^{~9}_{\Lambda\Lambda}$Li and $^{~9}_{\Lambda\Lambda}$Be are taken from 
four-body cluster model calculations in Ref.~\cite{Hi02}. The four pseudodata 
for $\Delta B_{\Lambda\Lambda}$ of heavier systems with $A\geq 18$ are obtained from the three-body 
model calculations presented in Appendix \ref{app:3BModel}.}
\begin{tabular}{cccc}
\hline\hline
Nuclide & $\Delta B_{\Lambda\Lambda}$ (MeV) & $B_{\Lambda\Lambda}$ (MeV) \\
\hline
$^{11}_{\Lambda\Lambda}$Be \cite{Ek19}  & --- & $\bf 19.07$ \\
$^{~9}_{\Lambda\Lambda}$Li \cite{Hi02}  & --- & $\bf 14.55^*$ \\
$^{~9}_{\Lambda\Lambda}$Be \cite{Hi02}  & --- & $\bf 14.40^*$ \\
$^{~18}_{\Lambda\Lambda}$O  & $\bf 1.22^*$ & $27.30^*$ \\
$^{~34}_{\Lambda\Lambda}$S  & $\bf 1.07^*$ & $36.42^*$ \\
$^{~42}_{\Lambda\Lambda}$Ca & $\bf 1.00^*$ & $39.03^*$ \\
$^{~58}_{\Lambda\Lambda}$Ni & $\bf 0.86^*$ & $42.60^*$ \\
\hline\hline
\end{tabular}
\label{tb:dataset}
\end{table}

\begin{figure}[t]
\includegraphics[width=\linewidth]{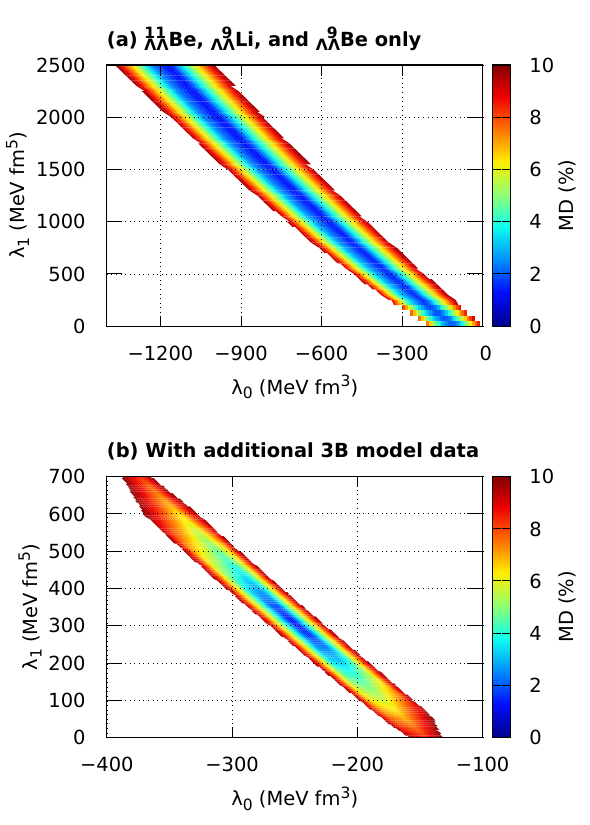}
\caption{MD values defined in Eq.~\eqref{eq:MD} as functions of the $\Lambda\Lambda$ 
interaction parameters $\lambda_0$ and $\lambda_1$ for (a) the restricted dataset consisting of 
$^{11}_{\Lambda\Lambda}$Be, $^{~9}_{\Lambda\Lambda}$Li, and $^{~9}_{\Lambda\Lambda}$Be, 
and (b) the full dataset listed in Table~\ref{tb:dataset}. 
The remaining parameters are fixed at $\lambda_2=0$ and $\lambda_3=0$.
KIDS0 and KIDS0-Y4 parameter sets are used for $NN$ and $N\Lambda$ interactions, respectively. }
\label{fig:MD_a0-a1}
\end{figure}

We perform two-parameter fittings with $\lambda_2=\lambda_3=0$. The parameter sets are labeled as 
``KIDS(0/-A/-B/-C/-D)-Y4-LL2'' or, if not confusing, ``LL2'' for short. 
Figure~\ref{fig:MD_a0-a1} shows the MD values [Eq.~\eqref{eq:MD}] as functions of the $\Lambda\Lambda$ 
interaction parameters $\lambda_0$ and $\lambda_1$ for (a) dataset restricted to $^{11}_{\Lambda\Lambda}$Be, 
$^{~9}_{\Lambda\Lambda}$Li, and $^{~9}_{\Lambda\Lambda}$Be and (b) the full set given in Tab.~\ref{tb:dataset}. 
Here, KIDS0 and KIDS0-Y4 parameter sets are used for $NN$ and $N\Lambda$ interactions, respectively.
As seen in Fig.~\ref{fig:MD_a0-a1}(a), the dataset restricted to the light systems
is insufficient to determine both parameters simultaneously. 
The MD landscape exhibits an elongated valley in the $(\lambda_0,\lambda_1)$ plane, 
effectively constraining only a quasi-linear combination of the two parameters.
In contrast, Fig.~\ref{fig:MD_a0-a1}(b) shows that 
the full dataset, which includes heavier systems, 
provides a well-defined minimum and thus enables a more reliable and simultaneous determination of both $\lambda_0$ and $\lambda_1$.
This confirms the preceding argument that a dataset covering a wide range of mass numbers 
is essential for simultaneously constraining the bulk and surface components of the functional. 
Table~\ref{tb:paramsets-LL2} summarizes the LL2 parameter sets obtained by minimizing 
the MD value for the full dataset.

\begin{table}
\caption{Parameter sets in the $\Lambda\Lambda$ sector obtained from fittings to full dataset 
in Table~\ref{tb:dataset} together with the corresponding MD values. 
The Y4 parameter sets are employed for the $N\Lambda$ sector~\cite{CHHC22}. 
The LL2 series, in which $\lambda_2=\lambda_3=0$, are listed.}
\begin{tabular}{cccccc}
\hline\hline
                          & KIDS0     & KIDS-A    & KIDS-B    & KIDS-C    & KIDS-D \\
                          & LL2       & LL2       & LL2       & LL2       & LL2    \\
\hline
$\lambda_0$ (MeV\,fm$^3$) & $-249.32$ & $-232.01$ & $-246.31$ & $-253.39$ & $-240.61$ \\
$\lambda_1$ (MeV\,fm$^5$) & $ 305.70$ & $ 268.19$ & $ 300.40$ & $ 314.53$ & $ 286.95$ \\
$\lambda_2$ (MeV\,fm$^5$) & $0$       & $0$       & $0$       & $0$       & $0$ \\
$\lambda_3$ (MeV\,fm$^6$) & $0$       & $0$       & $0$       & $0$       & $0$ \\
MD (\%)                   & $1.13$    & $2.80$    & $1.49$    & $0.89$    & $2.45$ \\
\hline\hline
\end{tabular}
\label{tb:paramsets-LL2}
\end{table}

\begin{widetext}

Table~\ref{tb:BLL-and-DelBLL} compares between data and 
calculated values for $B_{\Lambda\Lambda}$ and $\Delta B_{\Lambda\Lambda}$ 
for nuclei included in the dataset. 
The sets LL2 reproduce the dataset reasonably well, especially
the $\Delta B_{\Lambda\Lambda}$ pseudodata. 
This is reasonable given that the mean-field theory is suitable to heavier systems. 

\begin{table}
\caption{Comparison between data and calculated values for $B_{\Lambda\Lambda}$ 
and $\Delta B_{\Lambda\Lambda}$ in unit of MeV. 
}
\begin{tabular}{ccccccc}
\hline\hline
Observables & \ \ Data\ \                                                         & KIDS0   & KIDS-A  & KIDS-B  & KIDS-C  & KIDS-D\\
            &                                                                     & LL2     & LL2     & LL2     & LL2     & LL2   \\
\hline
$B_{\Lambda\Lambda}\left( ^{11}_{\Lambda\Lambda}{\rm Be}\right)$ & $19.07$        & $18.37$ & $17.78$ & $18.24$ & $18.44$ & $17.89$\\
$B_{\Lambda\Lambda}\left( ^{~9}_{\Lambda\Lambda}{\rm Li}\right)$ & $14.55$        & $14.45$ & $13.70$ & $14.22$ & $14.55$ & $13.87$\\
$B_{\Lambda\Lambda}\left( ^{~9}_{\Lambda\Lambda}{\rm Be}\right)$ & $14.40$        & $14.44$ & $13.67$ & $14.19$ & $14.53$ & $13.85$\\
$\Delta B_{\Lambda\Lambda}\left( ^{~18}_{\Lambda\Lambda}{\rm O }\right)$ & $1.22$ & $ 1.22$ & $ 1.22$ & $ 1.22$ & $ 1.22$ & $ 1.22$\\
$\Delta B_{\Lambda\Lambda}\left( ^{~34}_{\Lambda\Lambda}{\rm S }\right)$ & $1.07$ & $ 1.08$ & $ 1.07$ & $ 1.08$ & $ 1.07$ & $ 1.08$\\
$\Delta B_{\Lambda\Lambda}\left( ^{~42}_{\Lambda\Lambda}{\rm Ca}\right)$ & $1.00$ & $ 0.98$ & $ 0.99$ & $ 0.99$ & $ 0.99$ & $ 0.98$\\
$\Delta B_{\Lambda\Lambda}\left( ^{~58}_{\Lambda\Lambda}{\rm Ni}\right)$ & $0.86$ & $ 0.86$ & $ 0.87$ & $ 0.86$ & $ 0.87$ & $ 0.86$\\
\hline\hline
\end{tabular}
\label{tb:BLL-and-DelBLL}
\end{table}

\end{widetext}

We try also three-parameter ($\lambda_0$, $\lambda_1$, and $\lambda_3$) fits 
to the full dataset. These parameter sets are labeled ``KIDS(0/-A/-B/-C/-D)-Y4-LL3'' or ``LL3'' for short.
Table~\ref{tb:paramsets-LL3} summarizes the LL3 parameter sets obtained by minimizing 
the MD value for the full dataset. 
The inclusion of the $\lambda_3$ term does not lead to any significant improvement in the quality of the fit. 
We have also examined fits with $\lambda_3$ fixed at positive values 
and found that excessively large values of $\lambda_3$ deteriorate the agreement with the hypernuclear data, which constrains $\lambda_3$ only loosely (see next subsection). 

\begin{table}
\caption{LL3 Parameter sets in the $\Lambda\Lambda$ sector obtained from fittings to full dataset 
in Table~\ref{tb:dataset} together with the corresponding MD values. 
The Y4 parameter sets are employed for the $N\Lambda$ sector~\cite{CHHC22}.}
\begin{tabular}{cccccc}
\hline\hline
                          & KIDS0     & KIDS-A    & KIDS-B    & KIDS-C    & KIDS-D \\
                          & LL3       & LL3       & LL3       & LL3       & LL3    \\
\hline
$\lambda_0$ (MeV\,fm$^3$) & $-272.14$ & $-222.29$ & $-261.12$ & $-238.01$ & $-253.72$ \\
$\lambda_1$ (MeV\,fm$^5$) & $ 321.23$ & $ 257.48$ & $ 309.48$ & $ 295.96$ & $ 294.50$ \\
$\lambda_2$ (MeV\,fm$^5$) & $0$       & $0$       & $0$       & $0$       & $0$ \\
$\lambda_3$ (MeV\,fm$^6$) & $ 119.06$ & $ -35.82$ & $ 79.93$  & $-54.57$  & $ 70.19$ \\
MD (\%)                   & $0.96$    & $2.77$    & $1.35$    & $0.83$    & $2.32$ \\
\hline\hline
\end{tabular}
\label{tb:paramsets-LL3}
\end{table}

These results suggest that the density-dependent $\Lambda\Lambda$ interaction plays a secondary role in finite hypernuclei (at least for the systems included 
in the present dataset) but becomes relevant for dense matter. 
Additional information, such as constraints from neutron-star observables, is therefore required to further restrict $\lambda_3$ as well as $\lambda_2$, 
which is irrelevant to the ground states of hypernuclei.

\subsection{$(\lambda_2,\lambda_3)$ dependence of neutron-star properties}

In this section, we investigate the $(\lambda_2,\lambda_3)$ dependence of neutron-star properties within the present models. 
We explore the parameter space of $\lambda_2$ and $\lambda_3$ by requiring that the maximum mass of neutron stars exceeds $2M_\odot$~\cite{Demorest10,Arzoumanian18,Antoniadis13,Cromartie20,Fonseca21} 
and that the mass-radius (M-R) relation agrees with the NICER data for PSR~J0740+6620~\cite{Salmi24,Salmi24-data}. 
To this end, stellar configurations are obtained by solving the Tolman-Oppenheimer-Volkov (TOV) equation for a given EoS of neutron-star matter.

The EoSs for homogeneous $npe\mu$ or $np\Lambda e\mu$ matter in cold and neutrino-free neutron stars 
are constructed using the KIDS models under the conditions of charge neutrality and $\beta$ equilibrium~\cite{Glendenning}.
The crust EoSs for low densities are taken from Refs.~\cite{BPS,BBP}. 
The transition to the KIDS EoS for homogeneous matter is made at a mass density of 
$1.5\times 10^{14}$~g/cm$^3$, following Ref.~\cite{SLy230}.
To examine the causality of the EoSs, we evaluate the speed of sound, defined as 
\begin{align}
c_s^2 &= \frac{dP}{d\varepsilon}, 
\end{align}
where $P$ is the pressure given as a function of the energy density $\varepsilon$ obtained from an EoS. 
If the speed of sound exceeds the speed of light at any density reached in a neutron-star 
configuration obtained from the TOV equations, the corresponding stellar configuration is excluded from the M-R plot.

We focus on KIDS-A and KIDS-D models 
together with the corresponding hyperonic functionals.
Hereafter, we simultaneously include all the four parameters of ${\cal E}_{\Lambda\Lambda}$ 
in our consideration and label the parameter sets by ``KIDS-(A/D)-Y4-LL4($\lambda_2,\lambda_3$)'' or 
simply ``LL4($\lambda_2,\lambda_3$)'', 
where $\lambda_2$ and $\lambda_3$ are specified in units of MeV\,fm$^5$ and MeV\,fm$^6$, respectively. 

\begin{figure}
\includegraphics[width=\linewidth]{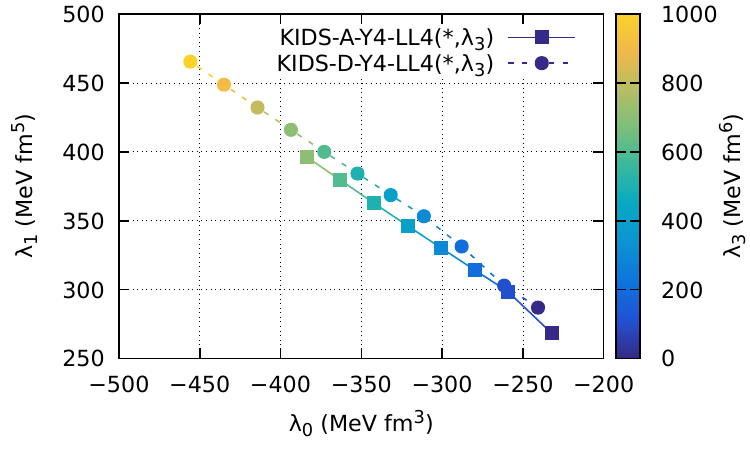}
\caption{Values of $(\lambda_0,\lambda_1)$ determined by fittings to hypernuclear data 
with $\lambda_3$ fixed at $0,\ 100,\ 200,\ \dots $~MeV\,fm$^6$.
The squares and circles represent the results for KIDS-A and -D, respectively, 
and the color indicates the value of $\lambda_3$.
The upper bounds of $\lambda_3$ is determined by the MD values. See main texts for detail.}
\label{fig:a0-a1}
\end{figure}

The LL4($\lambda_2,\lambda_3$) parameter sets are constructed as follows. 
For a given value of $\lambda_3$, the parameters $\lambda_0$ and $\lambda_1$ are refitted to the data listed in Tab.~\ref{tb:dataset} for the double-$\Lambda$ hypernuclei. 
We consider $\lambda_3=0,\ 100,\ 200,\ \dots$~MeV\,fm$^6$ and 
impose an upper bound on $\lambda_3$ by the resulting MD value, 
requiring that it does not exceed the minimum MD value for the LL3 set shown in Tab.~\ref{tb:paramsets-LL3} by more than 1\%. 
This practical criterion of the 1\% threshold is motivated by the fact that the pseudodata 
for $\Delta B_{\Lambda\Lambda}\sim 1$~MeV are specified to three significant digits. 
Nevertheless, this choice is to some extent arbitrary. 
The resulting upper bounds of $\lambda_3$ are found to be $\sim 700$~MeV\,fm$^6$ for KIDS-A and $\sim 1000$~MeV\,fm$^6$ for KIDS-D.
See Fig.~\ref{fig:a0-a1} for the optimized values of $(\lambda_0,\lambda_1)$ for different values of $\lambda_3$ 
ranging from $0$ to its upper bound. The squares and circles represent the results for KIDS-A-Y4-LL4($*,\lambda_3$) 
and KIDS-D-Y4-LL4($*,\lambda_3$), respectively, and the color indicates the value of $\lambda_3$.
(Tables of all parameter sets examined in this study are provided in the Supplemental Materials.)
As discussed in the previous subsection, $\lambda_2$ is irrelevant to the ground-state properties of 
double-$\Lambda$ hypernuclei. 
Accordingly, for a given value of $\lambda_3$ and a given KIDS model (A or D), 
the values of $\lambda_0$ and $\lambda_1$ are common to all values of $\lambda_2$.

\begin{figure}
\includegraphics[width=\linewidth]{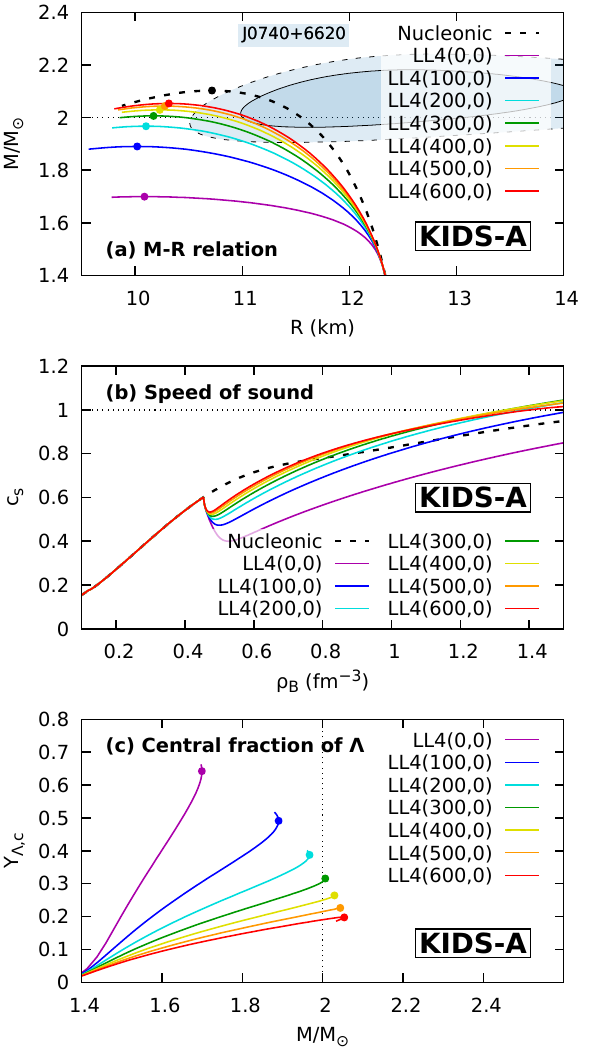}
\caption{(a) Mass-radius (M-R) relations of neutron stars 
calculated with different EoSs.
The black-dashed curve corresponds to nucleonic matter (KIDS-A), 
while the colored curves show the results for hyperonic matter 
based on the KIDS-A + KIDS-A-Y4 + KIDS-A-Y4-LL4($\lambda_2,0$) models.
Each curve is terminated at the central density where the causality condition 
is violated anywhere inside the star. 
Filled circles indicate the maximum-mass configurations 
when they are reached before the causality limit.
The light-blue regions enclosed by solid and dashed lines indicate 
the $1\sigma$ and $2\sigma$ credible regions, respectively, of NICER data 
for PSR~J0740+6620~\cite{Salmi24,Salmi24-data}.
(b) Speed of sound as a function of the baryon density for the corresponding EoSs.
(c) Fraction of $\Lambda$ hyperons at the centers of neutron stars as a function of stellar mass. 
}
\label{fig:MR-cs-YL_KIDSA-LL4-a2-0}
\end{figure}

\begin{figure}
\includegraphics[width=\linewidth]{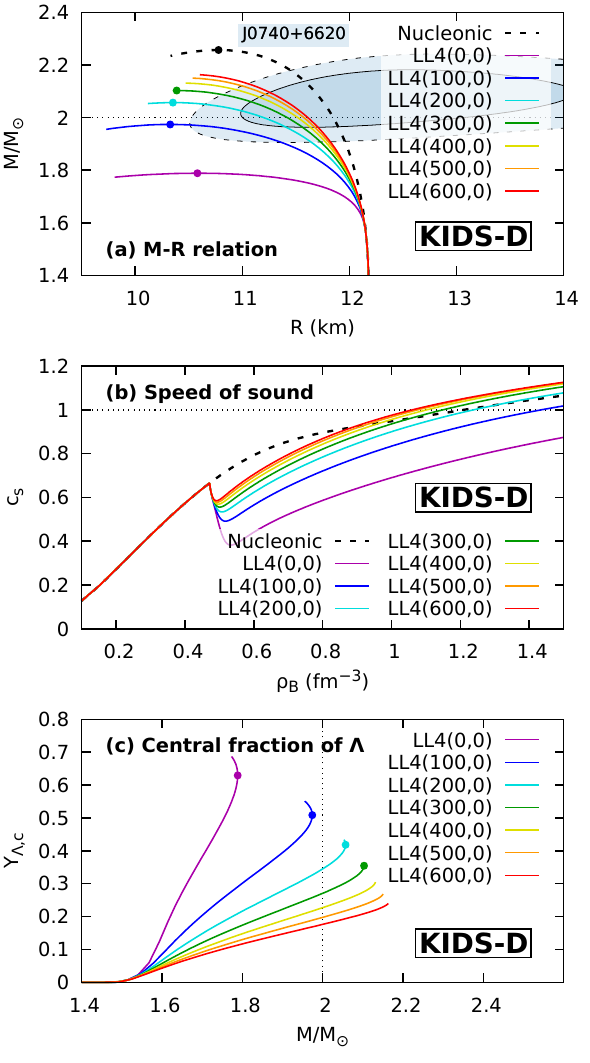}
\caption{Same as Fig.~\ref{fig:MR-cs-YL_KIDSA-LL4-a2-0} but for KIDS-D model.}
\label{fig:MR-cs-YL_KIDSD-LL4-a2-0}
\end{figure}

Figures~\ref{fig:MR-cs-YL_KIDSA-LL4-a2-0} and \ref{fig:MR-cs-YL_KIDSD-LL4-a2-0} show 
(a) the M-R relations, (b) the speed of sound as a function of the baryon density $\rho_B$, 
and (c) the fraction of $\Lambda$ hyperons at the centers of neutron stars 
as a function of stellar mass, 
which are obtained with EoSs of KIDS-A-Y4-LL4($\lambda_2,0$) and KIDS-D-Y4-LL4($\lambda_2,0$) models, 
respectively, with $\lambda_2$ ranging from 0 to 600~MeV\,fm$^5$. 
Here, the density-dependent term is turned off ($a_3=0$).  
In these figures, the black-dashed curves represent the results obtained with nucleonic-matter EoSs without $\Lambda$ hyperons, while the colored curves correspond to hyperonic-matter EoSs.
Each M-R curve is terminated at the point where the causality condition is violated anywhere inside the star. 
Filled circles indicate the maximum mass configurations when they are reached before the causality limit.
The light-blue regions enclosed by solid and dashed lines indicate 
the $1\sigma$ and $2\sigma$ credible regions, respectively, of NICER data for PSR~J0740+6620~\cite{Salmi24,Salmi24-data}.
For each of the KIDS-A and KIDS-D models, the corresponding Y4 parameter set is adopted for the $N\Lambda$ sector. 

As seen in Figs.~\ref{fig:MR-cs-YL_KIDSA-LL4-a2-0}(a) and \ref{fig:MR-cs-YL_KIDSD-LL4-a2-0}(a), 
for the both KIDS models without hyperons, the maximum neutron-star mass exceeds $2M_\odot$.
When hyperons are included with $\lambda_2 = 0$, however, the EoSs become significantly softer, 
and the maximum mass is reduced to values below $2M_\odot$, illustrating a so-called hyperon puzzle.
This behavior can be understood from the corresponding speed-of-sound curves shown in Figs.~\ref{fig:MR-cs-YL_KIDSA-LL4-a2-0}(b) and \ref{fig:MR-cs-YL_KIDSD-LL4-a2-0}(b).
As seen in the violet ($\lambda_2 = 0$) curves, the emergence of the new baryonic degree of 
freedom at densities $\rho_B \approx 0.4-0.5$~fm$^{-3}$ softens 
the EoS, and this softening persists up to high densities of $\rho_B \sim 1.5$~fm$^{-3}$.
For $\lambda_2 > 0$, the speed-of-sound curves indicate a recovery of stiffness due to 
the additional $\Lambda\Lambda$ repulsion.
Consequently, the maximum neutron-star mass exceeds $2M_\odot$ 
for $\lambda_2 \gtrsim 300$~MeV\,fm$^5$ in the KIDS-A models, 
and for $\lambda_2 \gtrsim 100$~MeV\,fm$^5$ in the KIDS-D models. 
For sufficiently large values of $\lambda_2$, the M-R curves also pass through the $2\sigma$ or $1\sigma$ credible regions of the NICER observation for PSR~J0740+6620.
However, the M-R curves tend to saturate as $\lambda_2$ increases, and 
the maximum mass as well converges to $(2.0-2.2)M_\odot$, 
making it difficult to infer an upper bound of $\lambda_2$ by the present 
neutron-star M-R observables alone. 
Note that the scan over $\lambda_2$ is similar to that performed by Togashi et al.~\cite{Togashi16}, 
who investigated the impact of the $p$-wave $\Lambda\Lambda$ interaction on neutron-star properties by varying its strength within a finite-range potential model. What is observed here is qualitatively 
similar to their results.

The relative fractions of $\Lambda$ hyperons at the centers of neutron stars, $Y_{\Lambda,c}$, 
are shown in Figs.~\ref{fig:MR-cs-YL_KIDSA-LL4-a2-0}(c) and \ref{fig:MR-cs-YL_KIDSD-LL4-a2-0}(c). 
For the maximum-mass neutron stars, we find
$Y_{\Lambda,c}\approx 0.6-0.7$ for $\lambda_2=0$, whereas 
$Y_{\Lambda,c}\approx 0.2-0.3$ for $\lambda_2=600$~MeV\,fm$^{5}$ in both models.
Within the present framework, the central $\Lambda$-hyperon fraction in 
massive neutron stars is typically found to be $Y_\Lambda \sim 0.2-0.3$.

\begin{figure}
\includegraphics[width=\linewidth]{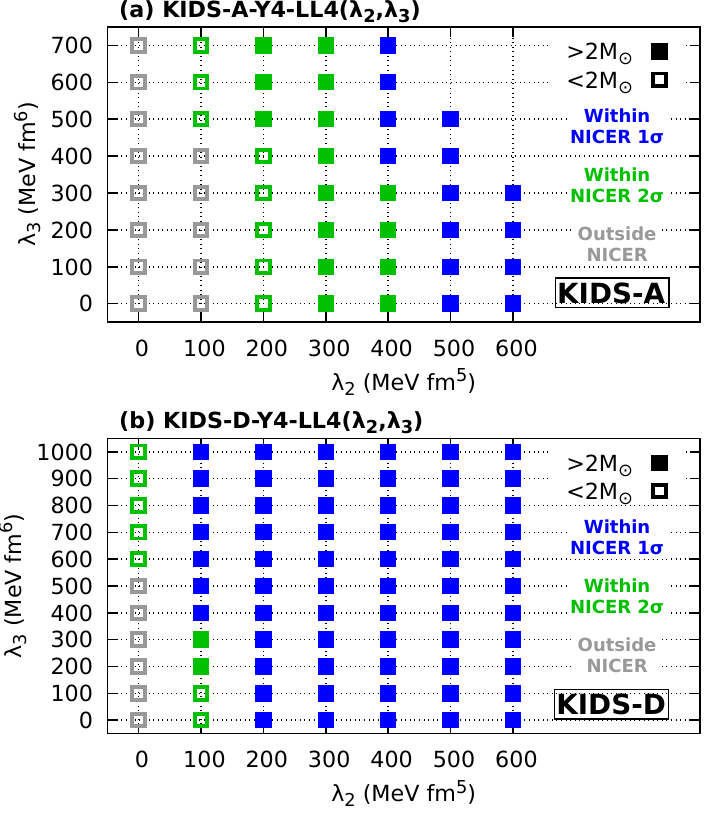}
\caption{Summary of the neutron-star properties in the $(\lambda_2,\lambda_3)$ plane for
(a) KIDS-A-Y4-LL4($\lambda_2,\lambda_3$) and 
(b) KIDS-D-Y4-LL4($\lambda_2,\lambda_3$) parameter sets.
Closed and open symbols indicate whether or not the maximum neutron-star mass exceeds $2M_\odot$. 
The consistency of the M-R curves with the NICER data for PSR~J0740+6620~\cite{Salmi24,Salmi24-data}
is distinguished by colors
(blue:~consistent within $1\sigma$; 
green:~consistent within $2\sigma$; 
gray:~outside the $2\sigma$ region).
}
\label{fig:test-a2-a3}
\end{figure}

Figure~\ref{fig:test-a2-a3} summarizes the neutron-star properties in the $(\lambda_2,\lambda_3)$ plane 
for (a) KIDS-A and (b) KIDS-D.
The figure indicates whether the neutron-star mass exceeds $2M_\odot$ and 
whether the M-R curve passes through the credible region of NICER data for PSR~J0740+6620. 
The colored and filled points in the $(\lambda_2,\lambda_3)$ plane may be regarded phenomenologically acceptable. 
As discussed earlier, the parameter $\lambda_3$ is bounded from above by the double-$\Lambda$ hypernuclear data, 
whereas an upper bound of $\lambda_2$ cannot be determined from the neutron-star observables alone 
in the present analyses.
Within the explored parameter domain, 
the conditions from the present neutron-star observables on the M-R relation 
for massive neutron stars are more sensitive to $\lambda_2$ than to $\lambda_3$. 
The repulsive effect of $\lambda_3>0$, below its upper bound, slightly shifts the M-R curves upward for $M\gtrsim 2M_\odot$, 
reducing the threshold value of $\lambda_2$ by at most $\sim 100$~MeV\,fm$^{5}$.

As a final remark, we compare the present results with those of Ref.~\cite{SHHL26}. 
From Figs.~\ref{fig:a0-a1} and \ref{fig:test-a2-a3}, the acceptable parameter ranges 
in the present analysis can be approximately summarized as 
$-500 \lesssim \lambda_0 \lesssim -200$~(MeV\,fm$^{3}$), 
$250 \lesssim \lambda_1 \lesssim 500$~(MeV\,fm$^{5}$), 
$100 \lesssim \lambda_2 \lesssim 600$~(MeV\,fm$^{5}$), and 
$0 \leq \lambda_3 \lesssim 1000$~(MeV\,fm$^{6}$), 
where we restrict the discussion to $\lambda_3 \ge 0$.
Comparison with the posterior distributions reported in Ref.~\cite{SHHL26} indicates that the allowed regions for $\lambda_0$ and $\lambda_1$ show marginal overlap. 
A remarkable agreement is found for the $p$-wave parameter $\lambda_2$, whose maximum-likelihood value in Ref.~\cite{SHHL26} falls within our acceptable range. 
This consistency can be attributed to the fact that $\lambda_2$ is constrained solely by neutron-star properties in both frameworks. 
The agreement for the density-dependent parameter $\lambda_3$ is less pronounced. 
These differences can be understood in terms of the different inputs adopted in the two analyses. 
In particular, the present study relies on hypernuclear information supplemented by pseudodata spanning a wider mass range, whereas Ref.~\cite{SHHL26} 
uses the potential depth inferred from a limited set of light double-$\Lambda$ hypernuclear data and 
also treats the exponent $\alpha$ of the density-dependent term as an additional free parameter. 
We note that Ref.~\cite{SHHL26} obtains informative posterior distributions within the prior ranges they adopted, 
indicating that their datasets provide meaningful constraints on the $\Lambda\Lambda$ interaction.
Taken together, the two studies underscore the complementary roles of 
hypernuclear and astrophysical information in constraining the $\Lambda\Lambda$ interaction.

\section{Conclusions and Outlook}\label{sec:Conclusions}
In this work, we have investigated the $\Lambda\Lambda$ interaction in finite nuclei and dense matter
within a Skyrme energy density functional framework based on the KIDS models.
A Skyrme-type $\Lambda\Lambda$ interaction including the standard $s$- and $p$-wave terms,
as well as a density-dependent contribution representing an effective $N\Lambda\Lambda$
three-body force, has been employed.
The primary goal of this study was to constrain the poorly known $\Lambda\Lambda$ interaction
by combining information from double-$\Lambda$ hypernuclei and astronomical observations of neutron stars.

Using the available experimental data for light double-$\Lambda$ hypernuclei together with
pseudodata obtained from core + $2\Lambda$ three-body model calculations,
we have shown that the inclusion of heavier systems is essential for a simultaneous and reliable
determination of the bulk and surface components of the $\Lambda\Lambda$ interaction.
In particular, the parameters associated with the $s$-wave terms, $\lambda_0$ and $\lambda_1$,
can be constrained within a relatively narrow range once the pseudodata for medium-mass hypernuclei are included.
The resulting parameter sets reproduce the available double-$\Lambda$ binding energies and
$\Lambda\Lambda$ correlation energies with good accuracy.

We have further explored the roles of the $p$-wave $\Lambda\Lambda$ interaction and the
density-dependent $N\Lambda\Lambda$ term by examining neutron-star properties.
The emergence of $\Lambda$ hyperons softens the EoS and reduces the maximum
neutron-star mass below $2M_\odot$ when these repulsive components are absent.
However, we have demonstrated that a moderate repulsion provided by the $p$-wave term,
and to a lesser extent by the density-dependent term, restores sufficient stiffness of the EoS, 
allowing the EoS to be consistent with the observed masses and radii of massive neutron stars,
including the NICER constraints for PSR~J0740+6620.
Within the phenomenologically acceptable parameter region, the central $\Lambda$-hyperon fraction in massive
neutron stars is typically found to be $Y_{\Lambda,c} \sim 0.2-0.3$.

The present results indicate that the KIDS-based Skyrme energy density functional,
augmented by a suitably constrained $\Lambda\Lambda$ interaction,
provides a coherent description of both multi-$\Lambda$ hypernuclei and hyperonic neutron-star matter.
The present analysis also highlights the importance of systematic measurements of double-$\Lambda$ hypernuclei over a wide mass range, which would provide direct constraints on the $\Lambda\Lambda$ interaction complementary to astrophysical observations. 
Future experimental data on heavier double-$\Lambda$ hypernuclei and improved astronomical
constraints on neutron-star observables will be crucial for further refining the $\Lambda\Lambda$
and $N\Lambda\Lambda$ interactions and for achieving a more definitive resolution of the hyperon puzzle.

The present $\Lambda\Lambda$ interaction can serve as a starting point 
for systematic Skyrme EDF studies of double- and multi-$\Lambda$ hypernuclei, 
including their ground-state properties~\cite{KMGR15,GBKM18,Guo22,ZhHiSa25,CuLeSc00} or the collective properties~\cite{MiHa12,MCH09,MiCh11}. 
Applying the present dataset within relativistic mean-field frameworks based on flavor SU(3) symmetry~\cite{Glendenning,Sch94,MCS13,MCKS25}, 
which include additional hyperons such as $\Sigma$ and $\Xi$,
would also be an interesting direction for future work.

\acknowledgments
We are grateful to Emiko Hiyama for providing us the numerical GEM code for three-body calculations and 
for valuable discussions on the $\Lambda\Lambda$ interaction and its impact on double-$\Lambda$ hypernuclei. 
Y.~T. thanks Tsuyoshi Miyatsu for helpful discussions on neutron-star equations of state and related observables. 
Y.~T. acknowledges support from the Basic Science Research Program of the National Research Foundation of Korea (NRF) 
under grants No.~RS-2024-00361003, RS-2024-00460031, and RS-2021-NR060129. 
C. H. H. is grateful to Hana Gil for the discussions
on the neutron-star mass-radius relation.
Work of C. H. H. was supported by the NRF research grant No. 2023R1A2C1003177.
Work of M. K. C. was supported by the NRF research grant No. RS-2025-16071941.

%\begin{widetext}

\appendix

\section{Three-body model}\label{app:3BModel}

To generate pseudodata for $\Delta B_{\Lambda\Lambda}$,
we employ a core + $2\Lambda$ three-body model for double-$\Lambda$ hypernuclei.
The Hamiltonian is given by 
\begin{align}
H_{\rm 3B} &= \frac{\bm P_c^2}{2M_c} 
+ \sum_{i=1,2}\left[ \frac{\bm p_i^2}{2m_\Lambda} + V_{c\Lambda}(|\bm r_i-\bm r_c|) \right]
\nonumber
\\
&\hspace{.4cm}
+v_{\Lambda\Lambda}(|\bm r_1-\bm r_2|) - T_{\rm CM}, 
\end{align}
where $M_c$ and $\bm P_c$ are the mass and the momentum of the core normal nucleus, $\bm p_i$ ($i=1,2$) is the momentum of the $i$th $\Lambda$ hyperon, 
$V_{c\Lambda}$ and $v_{\Lambda\Lambda}$ denote the core-$\Lambda$ and $\Lambda\Lambda$ interactions, 
and $T_{\rm CM}$ is the center-of-mass (CM) kinetic energy. 

The three-body Schr\"odinger equation is solved using the Gaussian expansion method (GEM) 
\cite{HiKiKa02,Hiyama12}. 
To describe the ground state with $J^\pi=0^+$, 
only $s$-wave components are taken into account for the relative motions defined by the Jacobi coordinates. The core is treated as a spinless particle.

\subsection{Core-Lambda potential}

The core-$\Lambda$ potential is determined so as to reproduce 
the empirical single-$\Lambda$ binding energies $B_\Lambda$. 
The binding energy is given by the solution of the Schr\"odinger equation for the relative motion, 
\begin{align}
\left[ -\frac{\hbar^2}{2\mu}\bm\nabla^2 + V_{c\Lambda}(r) \right]\varphi(\bm r) &= -B_\Lambda\varphi(\bm r), 
\end{align}
where $\mu = M_cm_\Lambda/(M_c+m_\Lambda)$ is the reduced mass. 
The core mass is taken as $M_c = (A-1)m_N$, with $m_N = (m_n+m_p)/2 = 938.918754$~MeV, and the mass of $\Lambda$ is taken to be $m_\Lambda = 1115.68$~MeV~\cite{pdg}.

\begin{figure}
\includegraphics[width=\linewidth]{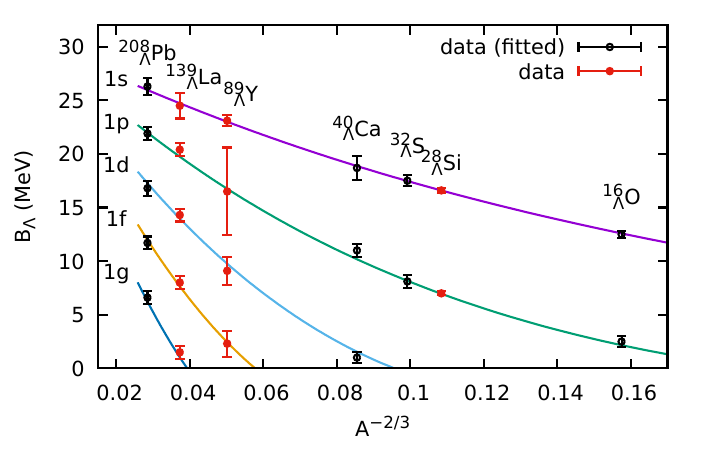}
\caption{Comparison of $B_\Lambda$'s between the WS-potential model and the data. 
The black symbols denote experimental data included in the fit, whereas 
the red symbols denote those not included in the fit.
}
\label{fig:B_L-systematic}
\end{figure}

We adopt a core-$\Lambda$ potential of a Woods-Saxon (WS) type: 
\begin{align}
V_{c\Lambda}(r)
&=
\frac{V_{c\Lambda}^0}{1+\exp{\left[ (r-R)/a \right]}}, \ \ \ R = r_0(A-1)^{1/3}. 
\end{align}
The three parameters $V_{c\Lambda}^0$, $r_0$, and $a$ are fitted to the measured 
$B_\Lambda$ values of $^{16}_\Lambda$O, $^{32}_\Lambda$S, and $^{40}_\Lambda$Ca, and $^{208}_\Lambda$Pb 
(see Table~\ref{tb:B_L-fitted}). 
The optimal parameters are listed in Table~\ref{tb:WS-cL}. 
A comparison between calculated and experimental $B_\Lambda$ values is shown in Fig.~\ref{fig:B_L-systematic}.

\begin{table}
\caption{The measured binding energies (MeV) of $\Lambda$ included in the fitting of WS parameters. 
The data are taken from Refs.~\cite{HaTa06,UsBo99}.}
\begin{tabular}{cccccc}
\hline\hline
Nuclide & $s$ orbit &  $p$ orbit &  $d$ orbit &  $f$ orbit &  $g$ orbit \\
\hline
$^{16}_\Lambda$O   & $12.5\pm 0.35$ & $2.5\pm 0.35$ & & &  \\
$^{32}_\Lambda$S   & $17.5\pm 0.5$  & $8.1\pm 0.6 $ & & &  \\
$^{40}_\Lambda$Ca  & $18.7\pm 1.1$  & $11.0\pm 0.6$ & $1.0\pm 0.5$ & & \\
$^{208}_\Lambda$Pb & $26.3\pm 0.8$   & $21.9\pm 0.6$ & $16.8\pm 0.7$ & $11.7\pm 0.6$ & $6.6\pm 0.6$ \\
\hline\hline
\end{tabular}
\label{tb:B_L-fitted}
\end{table}

\begin{table}
\caption{Optimal values of the WS parameters obtained by the fitting. }
\begin{tabular}{ccc}
\hline\hline
$V_{c\Lambda}^0$ (MeV) & $r_0$ (fm) & $a$ (fm) \\
\hline
$-30.1506$ & $1.1264$ & $0.6789$\\
\hline\hline
\end{tabular}
\label{tb:WS-cL}
\end{table}

\subsubsection{$^4$He-$\Lambda$ folding potential}\label{app:alpha-L}

The WS potential determined above is not suitable 
for the light system $^5_\Lambda$He and fails to reproduce the $B_\Lambda$ value of $3.12$~MeV~\cite{Juric73}. 
For $^5_\Lambda$He we instead use an $\alpha$-$\Lambda$ folding potential.

The $N\Lambda$ potential is taken as a one-range-Gaussian, 
\begin{align}
v_{N\Lambda}(r) &= v_{N\Lambda}^0\left( 1+\eta\bm\sigma_N\cdot\bm\sigma_\Lambda \right)e^{-r^2/\beta_{N\Lambda}^2}, 
\end{align}
with the parameters $\beta_{N\Lambda} = 1.034$~fm and $\eta = -0.1$~\cite{MBIY85}.
The strength $v_{N\Lambda}^0$ will be determined below.

The $\alpha$ particle is described as a $(1s)^4$ harmonic-oscillator configuration.
Its intrinsic density (after the center-of-mass correction) is given by
\begin{align}
\rho_\alpha(\bm r) &= 
4
\left( \frac{4\mu_\alpha}{3\pi} \right)^{3/2}
e^{-\frac{4}{3}\mu_\alpha r^2}, 
\end{align}
where $\mu_\alpha^{-1/2}$ is the oscillator length. 
The parameter $\mu_\alpha$ is adjusted so that the 
charge radius
\begin{align}
r_{\rm ch} &= \sqrt{\frac{9}{8\mu_\alpha}+0.841^2~{\rm fm}^2},
\end{align}
where the proton size of $0.841$~fm is quoted from PDG~\cite{pdg}, 
matches the measured value, $1.6755(28)$~fm~\cite{Angeli13}.
A value $1/\mu_\alpha = 1.366^2$~fm$^2$ yields $r_{\rm ch} = 1.6753$~fm. 

The $\alpha\Lambda$ folding potential is then given by~\cite{HKMYY97}
\begin{align}
V_{\alpha\Lambda}(r) 
&=
\int d^3r'\ \rho_\alpha(\bm r')v_{N\Lambda}(|\bm r-\bm r'|)
\nonumber
\\
%%%%%%%%%%%%%%%%%%%%%%%%%%%%%%%%%%%%%%%%%%%%%%%%%%%%%%%
&=
4v_{N\Lambda}^0 
\left( \frac{4\mu_\alpha}{4\mu_\alpha+3\mu_v} \right)^{3/2}
\exp\left( -
\frac{4\mu_\alpha\mu_v}{4\mu_\alpha+3\mu_v}
r^2 \right), 
\end{align}
where $\mu_v\equiv 1/\beta_{N\Lambda}^2$.
To reproduce $B_\Lambda=3.12$~MeV, $v_{N\Lambda}^0$ is readjusted to $-38.39$~MeV. 
The resulting parameters in $V_{\alpha\Lambda}$ are summarized in Table~\ref{tb:alpha-L}.

\begin{table}[]
\caption{Parameters of the $^4$He-$\Lambda$ folding potential. 
The oscillator length $\mu_\alpha$ is refitted to reproduce the measured charge radius of $^4$He. 
The strength $v_{N\Lambda}^0$ is refitted to reproduce $B_\Lambda\left( ^5_\Lambda{\rm He} \right) = 3.12$~MeV.}
\begin{tabular}{ccc}
\hline\hline
$\mu_\alpha^{-1/2}$ (fm) & $\beta_{N\Lambda}$ (fm) & $v_{N\Lambda}^0$ (MeV) \\
\hline
$1.366$ & $1.034$ & $-38.39$ \\
\hline\hline
\end{tabular}
\label{tb:alpha-L}
\end{table}

\subsection{Lambda-Lambda potential}

For the $\Lambda\Lambda$ interaction we use a three-range Gaussian potential 
of~\cite{Hi02},
\begin{align}
v_{\Lambda\Lambda}(r)
&=
\sum_{i=1}^3\left( v_i + v_i^\sigma\bm\sigma_\Lambda\cdot\bm\sigma_\Lambda\right)e^{-\mu_ir^2}. 
\end{align}
The parameters used in Ref.~\cite{Hi02} are listed in Table~\ref{tb:L-L}. 
To reproduce $B_{\Lambda\Lambda}=6.91$~MeV of NAGARA event~\cite{Ahn13,Takahashi01} 
using the present three-body model
(see App.~\ref{app:alpha-L}), the potential is multiplied by an overall renormalization factor,
\begin{align}
v_{\Lambda\Lambda} \to f_{\Lambda\Lambda}v_{\Lambda\Lambda},\ \ 
f_{\Lambda\Lambda} &= 0.9685. 
\end{align}

\begin{table}
\caption{Parameters of the $\Lambda\Lambda$ interaction taken from Ref.~\cite{Hi02}. In the present three-body model calculation, the overall factor of $f_{\Lambda\Lambda}=0.9685$ is applied to make a fine readjustment (see main texts). }
\begin{tabular}{cccc}
\hline\hline
$i$ & $\mu_i$ (fm$^{-2}$) & $v_i$ (MeV) & $v_i^\sigma$ (MeV) \\
\hline
1 & $0.555$ & $-10.67$ & $0.0966$ \\
2 & $1.656$ & $-93.51$ & $16.08$ \\
3 & $8.163$ & $4884$ & $915.8$ \\
\hline\hline
\end{tabular}
\label{tb:L-L}
\end{table}

Using the core-$\Lambda$ and the $\Lambda\Lambda$ interactions thus obtained, 
we perform three-body model calculations for several double hypernuclei. 
The resulting pseudodata for $B_{\Lambda\Lambda}$ and $\Delta B_{\Lambda\Lambda}$
are summarized in Table~\ref{tb:psdata}.
The values of $\Delta B_{\Lambda\Lambda}$ for
$^{18}_{\Lambda\Lambda}$O,
$^{34}_{\Lambda\Lambda}$S,
$^{42}_{\Lambda\Lambda}$Ca, and
$^{58}_{\Lambda\Lambda}$Ni
are used to constrain the $\Lambda\Lambda$ part of the functional in the main analysis.

\begin{table}
\caption{Calculated values of 
the $\Lambda\Lambda$ correlation energy $\Delta B_{\Lambda\Lambda}$, 
the double-$\Lambda$ binding energy $B_{\Lambda\Lambda}$, and 
the $\Lambda$ binding energy $B_\Lambda$ of the corresponding single-$\Lambda$ nucleus. 
Note that $\Delta B_{\Lambda\Lambda}(^{~A}_{\Lambda\Lambda}Z) 
= B_{\Lambda\Lambda}(^{~A}_{\Lambda\Lambda}Z)-2B_\Lambda(^{A-1}_{~~~\Lambda}Z)$.}
\begin{tabular}{cccc}
\hline\hline
Hypernucleus & $\Delta B_{\Lambda\Lambda}$ (MeV) & $B_{\Lambda\Lambda}$ (MeV) 
& $B_\Lambda\left({}^{A-1}_{~~~\Lambda} Z\right)$ (MeV) \\
\hline
$^{~~6}_{\Lambda\Lambda}$He & $0.67$ & $6.91$ & $3.12$ \\
$^{~18}_{\Lambda\Lambda}$O  & $1.22$ & $27.30$ & $13.04$ \\
$^{~34}_{\Lambda\Lambda}$S  & $1.07$ & $36.42$ & $17.67$ \\
$^{~42}_{\Lambda\Lambda}$Ca & $1.00$ & $39.03$ & $19.02$ \\
$^{~58}_{\Lambda\Lambda}$Ni & $0.86$ & $42.60$ & $20.87$ \\
\hline\hline
\end{tabular}
\label{tb:psdata}
\end{table}

\end{document}